# INVERSE RECONSTRUCTION OF MOVING CONTACT LOADS ON AN ELASTIC HALF-SPACE USING PRESCRIBLED SURFACE DISPLACEMENT

***Satoshi Takada[1]*, Yosuke Mori[2], Shintaro Hokada[2]***

*[1]Department of Mechanical Systems Engineering, Tokyo University of Agriculture and Technology,*
*2-24-16 Naka-cho, Koganei, Tokyo 184-8588, Japan*
*[2]Department of Industrial Technology and Innovation, Tokyo University of Agriculture and Technology,*
*2-24-16 Naka-cho, Koganei, Tokyo 184-8588, Japan*
**takada@go.tuat.ac.jp*

## Abstract

This study investigates the elastic response of a two-dimensional semi-infinite medium subjected to a moving surface load with a prescribed displacement profile. As a fundamental step, we derive analytical Green's functions for the displacement and stress fields generated by a point load traveling at a constant velocity along the surface, explicitly incorporating elastodynamic effects through Mach number dependence. These moving-load solutions serve as building blocks for constructing more general loading scenarios via linear superposition. Based on Green's functions, an inverse problem is formulated to reconstruct the unknown surface traction responsible for a given surface displacement. The inverse analysis is performed through a Fourier-domain inversion with regularization, which enables a direct and computationally efficient determination of the contact pressure without iterative forward simulations. This framework is applied to a rigid wheel–ground contact problem, where the imposed displacement is dictated by the wheel geometry. The reconstructed surface traction exhibits a smooth, symmetric distribution within the contact region, while the resulting subsurface stress fields are obtained in closed analytical form and involve dilogarithm functions. The principal stress difference reveals characteristic spatial patterns similar to photoelastic fringes, and their asymmetry increases with the Mach number, reflecting the dynamic nature of the moving contact.

## 1 Introduction

Contact mechanics is central to many engineering and geophysical problems, including tire–pavement interaction, soil–structure contact, and indentation testing [1]. A common modeling approach is to treat one body as an elastic half-space, which provides a tractable framework for analyzing stress and deformation fields due to external loads [1, 2].

Classical studies typically assume that the surface traction distribution is known. Fundamental solutions such as Boussinesq's (three dimensions) and Flamant's (two dimensions) describe the elastic response to a point load and serve as building blocks for analyzing more complex surface loadings via superposition [1]. These solutions have been widely applied to evaluate subsurface stress generated by prescribed surface loads [2, 3, 4]. Typical applications include wheel–ground and rail–wheel interactions, as well as off-road vehicle loading and soil–wheel contact problems [5].

In practice, however, the inverse situation often arises [6]: the surface displacement is known, while the corresponding traction is not. This occurs, for example, in wheel–ground contact, where the deformation profile is dictated by wheel geometry and ground compliance. In such cases, reconstructing the stress distribution responsible for a given displacement field constitutes a critical inverse problem.

This study addresses that problem through a two-step framework. First, we derive the displacement and stress fields generated by a point force acting on the surface of a two-dimensional elastic half-space under plane strain conditions. This solution forms the basis for constructing responses to arbitrary surface loads by linear superposition. Second, we formulate the inverse problem: given a prescribed surface displacement, we compute the traction distribution that would reproduce it. To ensure stability, the inverse formulation is solved numerically with regularization techniques.

Finally, we demonstrate the framework in a wheel–ground contact setting, where the imposed displacement corresponds to the wheel profile. The resulting surface and internal stress fields are computed and analyzed.

## 2 Displacement and stress distribution due to a point load

We consider a semi-infinite elastic medium characterized by the shear modulus $G$, mass density $\rho$, and Poisson's ratio $\nu$. A fixed coordinate system $(x', y', t')$ is introduced, where the elastic medium occupies the region $y' > 0$, and the surface is located at $y' = 0$. A concentrated load moving at a constant velocity $V$ in the negative $x'$-direction is applied on the surface. Here, the velocity $V$ is assumed to satisfy $V \ll v_{\mathrm{L}}, v_{\mathrm{T}}$, or equivalently, $M_{\mathrm{L}} \equiv V/v_{\mathrm{L}} \ll 1$ and $M_{\mathrm{T}} \equiv V/v_{\mathrm{T}} \ll 1$. This ensures a quasi-static dynamic regime, where

$$v_{\mathrm{L}} \equiv \sqrt{\frac{2(1-\nu)}{(1-2\nu)}\frac{G}{\rho}}, \qquad v_{\mathrm{T}} \equiv \sqrt{\frac{G}{\rho}} \tag{1}$$

are the longitudinal and transverse wave velocities of the elastic medium, respectively, and $M_{\mathrm{L}}$ and $M_{\mathrm{T}}$ denote the Mach numbers for the longitudinal and transverse waves, respectively.

Within the framework of elastodynamic theory, we analyze the stress field induced by the moving load. In the following, the state of plane strain is assumed. The motion of a linear elastic body, described by the displacement field $\boldsymbol{u}$, satisfies the elastodynamic equation

$$\rho\frac{\partial^2 \boldsymbol{u}}{\partial t^2} = G\left[\boldsymbol{\nabla}^2\boldsymbol{u} + \frac{1}{1-2\nu}\boldsymbol{\nabla}(\boldsymbol{\nabla}\cdot\boldsymbol{u})\right], \tag{2}$$

as given in Ref. [2]. According to Helmholtz's theorem, the displacement vector $\boldsymbol{u}$can be expressed in terms of a scalar potential $\phi$ and a vector potential $\boldsymbol{A}$ as [2]

$$\boldsymbol{u} = \boldsymbol{\nabla}\phi + \boldsymbol{\nabla}\times\boldsymbol{A}. \tag{3}$$

For the two-dimensional problem considered here, it is sufficient to take the vector potential as $\boldsymbol{A} = (0,0,A)^{\mathrm{T}}$. Since the focus of this study is on the effect of a load moving at a constant velocity, it is convenient to employ a coordinate system moving with the load, defined by $x = x' + Vt', y = y', t = t'$. By substituting Eq. (3) into the elastodynamic (Navier–Cauchy) equation (2) in the moving coordinate system, it is found that both potentials must satisfy the following wave equations:

$$\left(\frac{\partial^2}{\partial x^2} + \frac{\partial^2}{\partial y^2}\right)\phi = M_{\mathrm{L}}^2\frac{\partial^2\phi}{\partial x^2}, \tag{4a}$$

$$\left(\frac{\partial^2}{\partial x^2} + \frac{\partial^2}{\partial y^2}\right)A = M_{\mathrm{T}}^2\frac{\partial^2 A}{\partial x^2}. \tag{4b}$$

For later convenience, we introduce $\beta_{\mathrm{L}} \equiv \sqrt{1-M_{\mathrm{L}}^2}, \beta_{\mathrm{T}} \equiv \sqrt{1-M_{\mathrm{T}}^2}$.

In the moving coordinate system, the displacement and stress components are related to the potentials as follows:

$$u_x = \frac{\partial\phi}{\partial x} + \frac{\partial A}{\partial y}, \qquad u_y = \frac{\partial\phi}{\partial y} - \frac{\partial A}{\partial x}, \tag{5a}$$

$$\frac{\sigma_{xx}}{G} = (2 + M_{\mathrm{T}}^2 - 2M_{\mathrm{L}}^2)\frac{\partial^2\phi}{\partial x^2} + 2\frac{\partial^2 A}{\partial x\partial y}, \tag{5b}$$

$$\frac{\sigma_{yy}}{G} = (M_{\mathrm{T}}^2 - 2)\frac{\partial\phi}{\partial x^2} - 2\frac{\partial^2 A}{\partial x\partial y}, \tag{5c}$$

$$\frac{\sigma_{xy}}{G} = 2\frac{\partial^2\phi}{\partial x\partial y} + (M_{\mathrm{T}}^2 - 2)\frac{\partial^2 A}{\partial x^2}. \tag{5d}$$

We now introduce the Fourier transforms of the $x$-dependent parts of the two potentials as

$$\begin{Bmatrix}\hat{\phi}(k,y)\\ \hat{A}(k,y)\end{Bmatrix} = \frac{1}{\sqrt{2\pi}}\int_{-\infty}^{\infty}\begin{Bmatrix}\phi(x,y)\\ A(x,y)\end{Bmatrix}\mathrm{e}^{\mathrm{i}kx}\mathrm{d}x. \tag{6}$$

Substituting Eqs. (6) into the wave equations, the general solutions can be written as

$$\hat{\phi}(k,y) = f(k)\mathrm{e}^{-\beta_{\mathrm{L}}|k|y}, \qquad \hat{A}(k,y) = g(k)\mathrm{e}^{-\beta_{\mathrm{T}}|k|y}, \tag{7}$$

where $f(k)$ and $g(k)$ are unknown coefficients to be determined from the boundary conditions.

Similarly, by introducing the Fourier-transformed quantities of displacement and stress, denoted by $\hat{u}_\alpha$ and $\hat{\sigma}_{\alpha\beta}$, we obtain

$$\hat{u}_x = -\mathrm{i}kf(k)\mathrm{e}^{-\beta_{\mathrm{L}}|k|y} - \beta_{\mathrm{T}}|k|g(k)\mathrm{e}^{-\beta_{\mathrm{T}}|k|y}, \tag{8a}$$

$$\hat{u}_y = -\beta_{\mathrm{T}}|k|f(k)e^{-\beta_{\mathrm{L}}|k|y} + \mathrm{i}kg(k)\mathrm{e}^{-\beta_{\mathrm{T}}|k|y}, \tag{8b}$$

$$\begin{aligned}\frac{\hat{\sigma}_{xx}}{G} &= -(2 + M_{\mathrm{T}}^2 - 2M_{\mathrm{L}}^2)k^2 f(k)\mathrm{e}^{-\beta_{\mathrm{L}}|k|y}\\ &\quad +2\mathrm{i}\beta_{\mathrm{T}}k|k|g(k)\mathrm{e}^{-\beta_{\mathrm{T}}|k|y},\end{aligned} \tag{8c}$$

$$\begin{aligned}\frac{\hat{\sigma}_{yy}}{G} &= -(M_{\mathrm{T}}^2 - 2)k^2 f(k)\mathrm{e}^{-\beta_{\mathrm{L}}|k|y}\\ &\quad -2\mathrm{i}\beta_{\mathrm{T}}k|k|g(k)\mathrm{e}^{-\beta_{\mathrm{T}}|k|y},\end{aligned} \tag{8d}$$

$$\begin{aligned}\frac{\hat{\sigma}_{xy}}{G} &= 2\mathrm{i}\beta_{\mathrm{L}}k|k|f(k)\mathrm{e}^{-\beta_{\mathrm{L}}|k|y}\\ &\quad -(M_{\mathrm{T}}^2 - 2)k^2 g(k)\mathrm{e}^{-\beta_{\mathrm{T}}|k|y}.\end{aligned} \tag{8e}$$

To derive the Green's function, we first consider the case where a unit normal stress acts on the surface at $x = x_0$. The boundary conditions are given by

$$\sigma_{yy}|_{y=0} = -\delta(x - x_0), \qquad \sigma_{xy}|_{y=0} = 0, \tag{9}$$

where $\delta(x - x_0)$ denotes the Dirac delta function. By requiring that the expressions obtained from Eq. (8) at $y = 0$ satisfy the boundary conditions (9) in the Fourier domain, the coefficients $f(k)$ and $g(k)$ are determined as

$$f(k) = \frac{M_{\mathrm{T}}^2 - 2}{\sqrt{2\pi}G\mathcal{D}}\frac{\mathrm{e}^{\mathrm{i}kx_0}}{k^2}, \qquad g(k) = \frac{2\mathrm{i}\beta_{\mathrm{L}}}{\sqrt{2\pi}G\mathcal{D}}\frac{|k|\mathrm{e}^{\mathrm{i}kx_0}}{k^3}, \tag{10}$$

where

$$\mathcal{D} \equiv (M_{\mathrm{T}}^2 - 2)^2 - 4\beta_{\mathrm{L}}\beta_{\mathrm{T}}. \tag{11}$$

Performing the inverse Fourier transforms, the displacement and stress components can be expressed as

$$u_\alpha(x,y) = \mathcal{G}_\alpha^{(1)}(x - x_0, y), \sigma_{\alpha\beta}(x,y) = \mathcal{G}_{\alpha\beta}^{(2)}(x - x_0, y), \quad (12)$$

where $\mathcal{G}_\alpha^{(1)}$ and $\mathcal{G}_{\alpha\beta}^{(2)}$ are the Green's functions for displacement and stress, respectively. These are given by

$$\mathcal{G}_x^{(1)}(x,y) \equiv \frac{1}{\pi G \mathcal{D}} \left[ \begin{array}{c} (1+\beta_\mathrm{T}^2)\tan^{-1}\left(\frac{x}{\beta_\mathrm{L} y}\right) \\ -2\beta_\mathrm{L}\beta_\mathrm{T}\tan^{-1}\left(\frac{x}{\beta_\mathrm{T} y}\right) \end{array} \right], \quad (13a)$$

$$\mathcal{G}_y^{(1)}(x,y) \equiv -\frac{1}{\pi G \mathcal{D}} \left[ \begin{array}{c} \beta_\mathrm{T}(1+\beta_\mathrm{T}^2)\log\sqrt{x^2+\beta_\mathrm{L}^2 y^2} \\ -2\beta_\mathrm{L}\log\sqrt{x^2+\beta_\mathrm{T}^2 y^2} \end{array} \right], \quad (13b)$$

$$\mathcal{G}_{xx}^{(2)}(x,y) \equiv \frac{1}{\pi \mathcal{D}} \left[ \begin{array}{c} (1-\beta_\mathrm{T}^2+2\beta_\mathrm{L}^2)(1+\beta_\mathrm{T}^2)\frac{\beta_\mathrm{L} y}{x^2+\beta_\mathrm{L}^2 y^2} \\ -4\beta_\mathrm{L}\beta_\mathrm{T}\frac{\beta_\mathrm{T} y}{x^2+\beta_\mathrm{T}^2 y^2} \end{array} \right], \quad (13c)$$

$$\mathcal{G}_{yy}^{(2)}(x,y) \equiv -\frac{1}{\pi \mathcal{D}} \left[ \begin{array}{c} (1+\beta_\mathrm{T}^2)^2\frac{\beta_\mathrm{L} y}{x^2+\beta_\mathrm{L}^2 y^2} \\ -4\beta_\mathrm{L}\beta_\mathrm{T}\frac{\beta_\mathrm{T} y}{x^2+\beta_\mathrm{T}^2 y^2} \end{array} \right], \quad (13d)$$

$$\begin{aligned} \mathcal{G}_{xy}^{(2)}(x,y) \equiv & -\frac{1}{\pi \mathcal{D}} 2\beta_\mathrm{L}(1+\beta_\mathrm{T}^2) \\ & \times \left[\frac{x}{x^2+\beta_\mathrm{L}^2 y^2} - \frac{x}{x^2+\beta_\mathrm{T}^2 y^2}\right]. \end{aligned} \quad (13e)$$

Using these Green's functions, the displacement and stress fields in the elastic half-space induced by an arbitrary distributed normal surface load $\sigma_{yy}^{\mathrm{ex}}(x)$ can be obtained by superposition as

$$u_\alpha(x,y) = \int \mathcal{G}_\alpha^{(1)}(x - x_0, y)\sigma_{yy}^{\mathrm{ex}}(x_0)\mathrm{d}x_0, \quad (14a)$$

$$\sigma_{\alpha\beta}(x,y) = \int \mathcal{G}_{\alpha\beta}^{(2)}(x - x_0, y)\sigma_{yy}^{\mathrm{ex}}(x_0)\mathrm{d}x_0. \quad (14b)$$

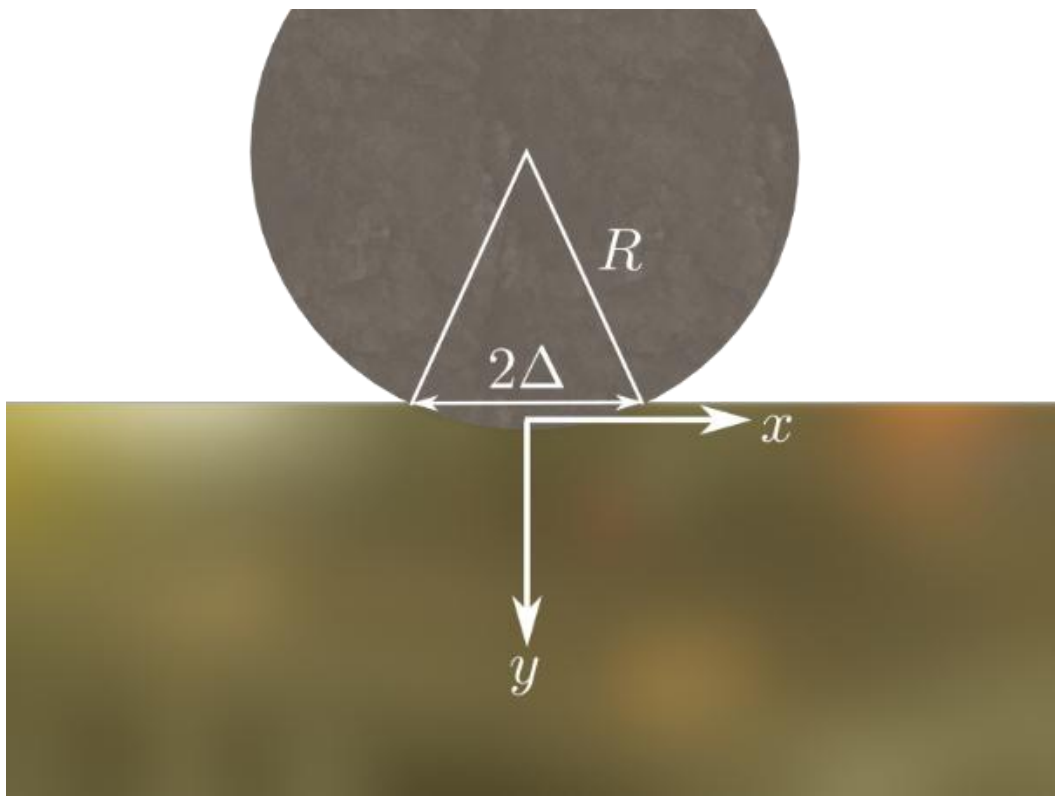

Fig. 1 Schematic picture of the analysis.

## 3 Inverse problem analysis

As shown in Fig. 1, let us consider a rigid wheel of radius $R$ that is indented only within the range $-\Delta \le x \le \Delta$. As in the previous sections, the wheel is assumed to move leftward with a constant velocity $V$. The deformation due to indentation is restricted to the $y$-direction displacement $u_y^w(x)$. Assuming that both the indentation depth and $\Delta$ are sufficiently small, the displacement can be expressed as

$$u_y^w(x) = \sqrt{R^2 - x^2} - \sqrt{R^2 - \Delta^2} \simeq \frac{\Delta^2 - x^2}{2R}. \quad \mathbf{(15)}$$

When the displacement given by Eq. (15) is produced on the surface due to an unknown normal stress $\sigma_{yy}^{\mathrm{ex}}(x)$, the compressive load can be determined by means of inverse problem analysis. By taking the Fourier transform of the governing equation, we obtain

$$\hat{u}_y(k,y) = \sqrt{2\pi}\hat{\mathcal{G}}(k,y)\hat{\sigma}_{yy}^{\mathrm{ex}}(k), \quad (16)$$

where $\hat{\sigma}_{yy}^{\mathrm{ex}}(k)$ is the Fourier transform of the unknown stress distribution $\sigma_{yy}^{\mathrm{ex}}(x)$, and $\hat{G}(k,y)$ represents the Fourier transform of the Green's function. The Fourier transform of the indentation displacement $u_y^w(x)$ is given by

$$\hat{u}_y^w(k) = \frac{1}{\sqrt{2\pi}}\frac{2}{R}\frac{\sin(\Delta k) - \Delta k\cos(\Delta k)}{k^3}. \quad (17)$$

Using Eq. (17) in Eq. (16) and solving for $\sigma_{yy}^{\mathrm{ex}}(x)$, the unknown load distribution can be obtained by performing the inverse Fourier transform:

$$\begin{aligned} \sigma_{yy}^{\mathrm{ex}}(x) &= \int_{-\infty}^{\infty}\hat{\sigma}_{yy}^{\mathrm{ex}}(k)\mathrm{e}^{-\mathrm{i}kx}\mathrm{d}k \\ &= \frac{\mathcal{D}}{\beta_\mathrm{T}(1+\beta_\mathrm{T}^2) - 2\beta_\mathrm{L}}\frac{2G\Delta}{\pi R}\left[1 - \frac{x}{\Delta}\tanh^{-1}\left(\frac{x}{\Delta}\right)\right]. \end{aligned} \quad (18)$$

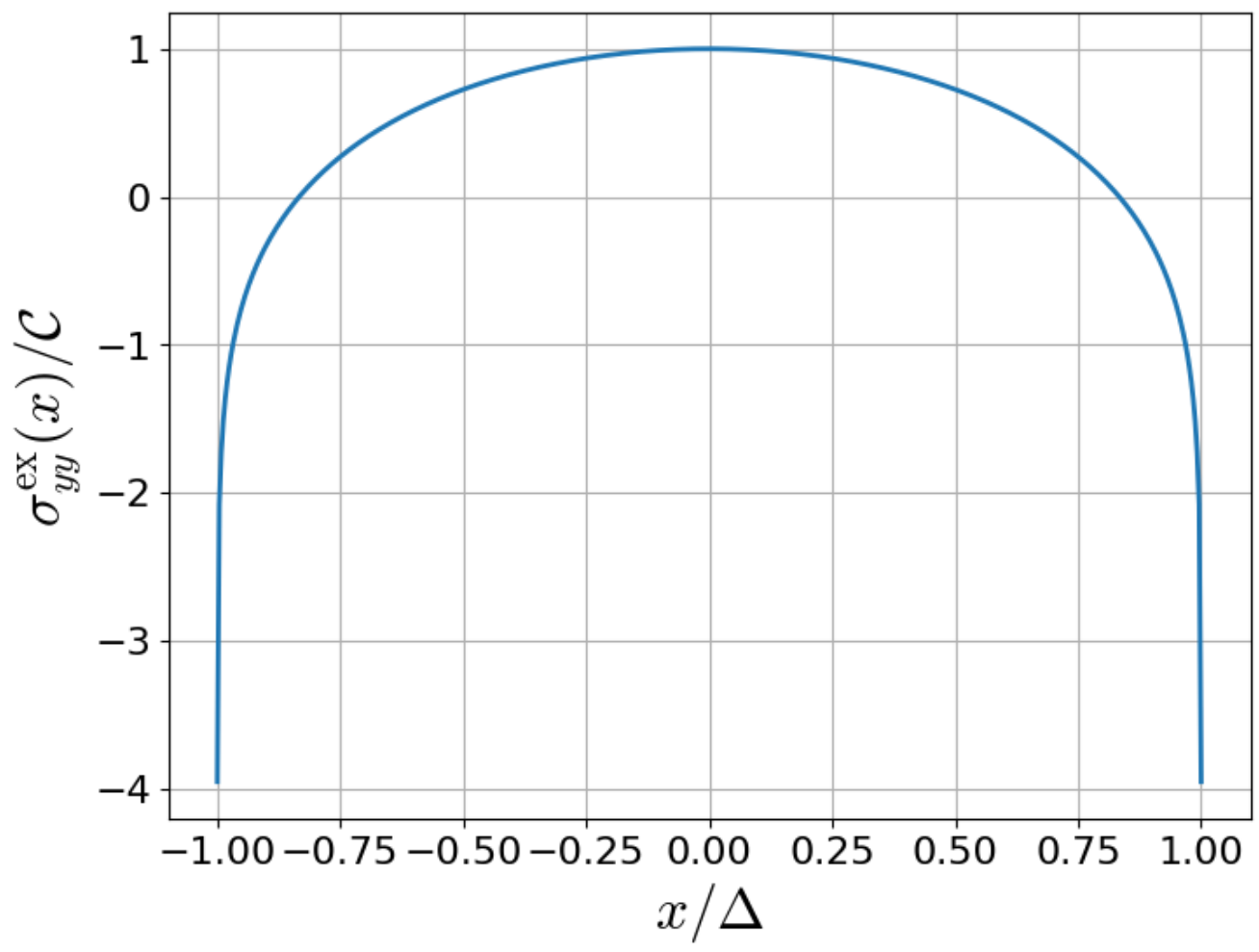

Fig. 2 Profile of $\sigma_{yy}^{\mathrm{ex}}(x)$ for $\nu = 0.3$ and $M_\mathrm{L} = 0.3$. The value is normalized by the characteristic length $\Delta$ and the shear modulus $G$.

Figure 2 shows the result obtained from the above analysis. For simplicity of notation, we introduce the constant

$$\mathcal{C} \equiv \frac{\mathcal{D}}{\beta_{\mathrm{T}}(1+\beta_{\mathrm{T}}^2)-2\beta_{\mathrm{L}}} \frac{2G\Delta}{\pi R}. \tag{19}$$

The reconstructed surface traction shown in Fig. 2 exhibits a symmetric distribution within the contact region, decreasing smoothly toward the edges at $x = \pm\Delta$. This feature is consistent with the expected behavior of a rigid wheel contacting an elastic half-space under small indentation, where the compressive stress is concentrated near the center of contact and gradually vanishes at the boundaries.

## 4 Stress distribution

The stress distribution induced by the load obtained in the previous section is now evaluated. Since the stress field obtained from Eq. (14b) can be expressed as a superposition, as given in Eq. (18), each stress component is calculated as follows:

$$\frac{\pi\mathcal{D}}{\mathcal{C}}\sigma_{xx}(x,y) = -(1-\beta_{\mathrm{T}}^2+2\beta_{\mathrm{L}}^2)(1+\beta_{\mathrm{T}}^2)\Im[\Phi(z_{\mathrm{L}})] + 4\beta_{\mathrm{L}}\beta_{\mathrm{T}}\Im[\Phi(z_{\mathrm{T}})], \tag{20a}$$

$$\frac{\pi\mathcal{D}}{\mathcal{C}}\sigma_{yy}(x,y) = (1+\beta_{\mathrm{T}}^2)^2\Im[\Phi(z_{\mathrm{L}})] - 4\beta_{\mathrm{L}}\beta_{\mathrm{T}}\Im[\Phi(z_{\mathrm{T}})], \tag{20b}$$

$$\frac{\pi\mathcal{D}}{\mathcal{C}}\sigma_{xy}(x,y) = -2\beta_{\mathrm{L}}(1+\beta_{\mathrm{T}}^2)\Re[\Phi(z_{\mathrm{L}})-\Phi(z_{\mathrm{T}})], \tag{20c}$$

where

$$\Phi(z) \equiv \log\frac{z+\Delta}{z-\Delta} + \frac{z}{2\Delta}\left[\mathrm{Li}_2\left(\frac{2\Delta}{\Delta+z}\right) + \mathrm{Li}_2\left(\frac{2\Delta}{\Delta-z}\right)\right], \tag{21}$$

and, for notational simplicity, we define

$$z_{\mathrm{L}} \equiv x + \mathrm{i}\beta_{\mathrm{L}}y, \qquad z_{\mathrm{T}} \equiv x + \mathrm{i}\beta_{\mathrm{T}}y. \tag{22}$$

Here, $\mathrm{Li}_2(z)$ denotes the dilogarithm (polylogarithm of order 2) [7]. In the numerical evaluations, the dilogarithm is computed using standard built-in functions in Python or other languages, and no Gibbs-type oscillations are observed in the resulting stress fields.

In photoelastic experiments, interference fringes appear as contours of the principal stress difference [8, 9, 10]

$$\sigma_1 - \sigma_2 \equiv \sqrt{\left(\sigma_{xx}-\sigma_{yy}\right)^2 + 4\sigma_{xy}^2}. \tag{23}$$

The analytical form of Eq. (18) reproduces this behavior accurately, providing a physically reasonable contact pressure profile. The subsurface stress field obtained from Eqs. (20) (see Fig. 3) reveals that the principal stress difference $\sigma_1 - \sigma_2$ forms characteristic lobes beneath the loaded region. These patterns qualitatively agree with the fringe distributions observed in photoelastic experiments for rolling contact. In particular, the shear-dominated zones appearing slightly ahead of and behind the contact center indicate the directional nature of the moving load. This asymmetry becomes more pronounced as the Mach number $M_{\mathrm{L}}$ increases, reflecting the dynamic amplification of stress propagation in the moving coordinate system.

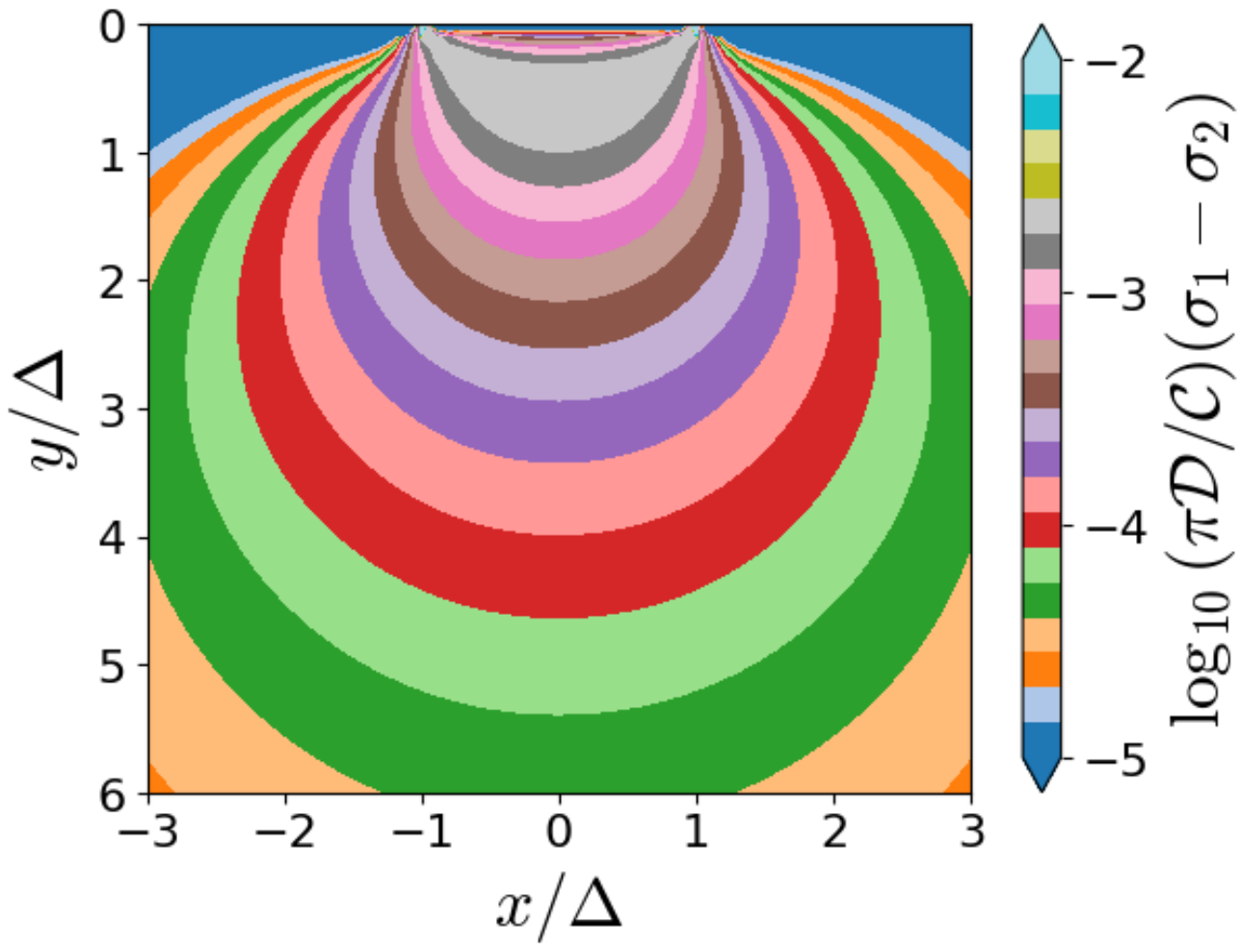

Fig. 3 Plot of the principal stress difference $\sigma_1 - \sigma_2$ for $\nu = 0.3$ and $M_{\mathrm{L}} = 0.3$. Here, the maxima appear at $x = \pm\Delta, y = 0$. The value is normalized by the characteristic length $\Delta$ and the shear modulus $G$.

## 5 Discussion

From the viewpoint of inverse analysis, it is important to emphasize that the present problem differs fundamentally from general inverse problems in elasticity and elastodynamics. In many conventional inverse formulations, the forward problem governed by partial differential equations must be solved repeatedly, often in conjunction with adjoint-based gradient evaluations. Moreover, the selection of appropriate regularization parameters typically requires extensive parameter searches, making the computational cost a dominant issue.

In contrast, the inverse problem considered here is computationally inexpensive. Because the Green's functions for a moving point load are available in closed analytical form, the forward response does not require numerical solution of boundary value problems. The reconstruction of the surface traction reduces to an algebraic division in the Fourier domain followed by an inverse transform, without any iterative calls to a forward solver. This feature significantly lowers the numerical cost compared with general inverse contact analyses and makes the proposed method well suited for parametric studies and benchmark testing of numerical models such as FEM or DEM simulations.

The present formulation captures both the static-like contact response at low velocities ($M_{\mathrm{L}}, M_{\mathrm{T}} \ll 1$) and the dynamic modification introduced by finite motion. Because the Mach numbers enter explicitly through $\beta_{\mathrm{L}}$ and $\beta_{\mathrm{T}}$, the model can continuously bridge between quasi-static and dynamic regimes.

Such analytical transparency provides a valuable benchmark for validating numerical elastodynamic solvers or discrete element simulations of rolling contact.

## 6 Conclusion

This study developed an analytical framework for evaluating the elastic response of a semi-infinite medium subjected to a moving surface load with a prescribed displacement profile. Starting from the elastodynamic equations, we derived the displacement and stress fields generated by a moving point load and constructed the corresponding Green's functions. By employing these solutions within an inverse formulation, the surface traction distribution required to reproduce an imposed displacement was obtained in closed form. The approach was demonstrated for the case of a rigid wheel moving at a constant velocity over an elastic half-space. The resulting load distribution and internal stress fields were expressed analytically, and the spatial variation of the principal stress difference was visualized. The analysis shows that the proposed inverse method effectively reconstructs realistic moving-contact stresses while retaining analytical transparency.

The present formulation provides a foundation for more general studies of dynamic contact problems, including rolling or sliding interactions, layered media, and viscoelastic effects. Future work will extend the method to three-dimensional geometries and incorporate experimental validation using photoelastic or numerical simulations.

The present analysis is subject to several limitations, including the assumptions of two-dimensional geometry, linear elasticity, and small indentation. Nevertheless, within these assumptions, the proposed inverse formulation provides an efficient and analytically transparent approach to reconstructing moving contact loads.

A notable advantage of the method is its very low computational cost. Once the Green's functions are obtained, the inverse reconstruction requires only spectral operations and no iterative forward simulations. This efficiency, combined with the availability of closed-form stress expressions, makes the present framework a valuable reference solution for both numerical and experimental studies of moving contact problems.

## 7 Acknowledgements

This work is partially supported by the Grant-in-Aid of MEXT for Scientific Research (Grant No. JP24K06974, No. JP24K07193, and No. JP25K01063).